\documentclass[conference]{IEEEtran}
\usepackage{cite}

\usepackage[pdftex]{graphicx}
\usepackage{subfig}
\usepackage{smartdiagram}
\usepackage{forest}
\usepackage{amsmath}
\usepackage{dirtree}
\usepackage{listings}
\usepackage{hyperref}
\usepackage{url}
\usepackage{outlines}

\begin{document}

\title{Modeling pre-Exascale AMR Parallel I/O Workloads via Proxy Applications}

\author{
\IEEEauthorblockN{William F Godoy}
\IEEEauthorblockA{
\textit{Computer Science and}\\
\textit{Mathematics Division,}\\
\textit{Oak Ridge National Laboratory}\\
Oak Ridge, TN, USA\\
Email: godoywf@ornl.gov}
\and
\IEEEauthorblockN{Jenna Delozier}
\IEEEauthorblockA{
\textit{College of Computing,}\\
\textit{Georgia Institute of Technology}\\
Atlanta, GA, USA\\
Email: jdelozier6@gatech.edu}
\and
\IEEEauthorblockN{Gregory R Watson}
\IEEEauthorblockA{
\textit{Computer Science and}\\
\textit{Mathematics Division,}\\
\textit{Oak Ridge National Laboratory}\\
Oak Ridge, TN, USA\\
Email: watsongr@ornl.gov}
}
\maketitle

\begin{abstract}

The present work investigates the modeling of pre-exascale input/output (I/O) workloads of Adaptive Mesh Refinement (AMR) simulations through a simple proxy application.
We collect data from the AMReX Castro framework running on the Summit supercomputer for a wide range of scales and mesh partitions for the hydrodynamic Sedov case as a baseline to provide sufficient coverage to the formulated proxy model. The non-linear analysis data production rates are quantified as a function of a set of input parameters such as output frequency, grid size, number of levels, and the Courant–Friedrichs–Lewy (CFL) condition number for each rank, mesh level and simulation time step.
Linear regression is then applied to formulate a simple analytical model which allows to translate AMReX inputs into MACSio proxy I/O application parameters, resulting in a simple ``kernel" approximation for data production at each time step. 
Results show that MACSio can simulate actual AMReX non-linear ``static" I/O workloads to a certain degree of confidence on the Summit supercomputer using the present methodology.
The goal is to provide an initial level of understanding of AMR I/O workloads via lightweight proxy applications models to facilitate autotune data management strategies in anticipation of exascale systems.
\end{abstract}



\IEEEpeerreviewmaketitle

\section{Introduction}
As we approach the exascale era in the next generation of supercomputers~\cite{8528398}, it is crucial to understand high-performance computing (HPC) computational, communication and input output (I/O) workloads in massive, large-scale scientific applications in order to effectively utilize the power of these systems~\cite{6375530,6495837}. This understanding becomes extremely complex and sophisticated due to the large number of factors involved in the end-to-end operation, from the existing algorithms in a scientific application or workflow, to the computing and communication patterns when fine-tuning the software stack parameters for a particular platform.

Adaptive Mesh Refinement (AMR)~\cite{BERGER1984484} is a powerful technique used for solving partial differential equations. AMR simulations running at scale are computationally and I/O intensive~\cite{1137736}. The I/O characteristics rely heavily on several aspects related to the parallelization, load balancing and domain decomposition strategies, user input specifying output frequency, and the problem-dependant configuration of the physics in a particular application. In addition, parallel I/O capabilities are often managed independently from the computational capabilities in large leadership facilities, thus applications can increasingly become I/O bound due to the different rates of improvements in hardware (parallel file system and storage throughput), and software (parallel I/O libraries) when compared to computational capacity~\cite{1137736}.

Proxy applications, also known as miniapps, are lightweight, relatively simple programs that enable replication of key characteristics of full-scale applications to facilitate the process of application optimization at scale~\cite{Heroux09}. They enable the abstraction of the computational aspects that drive performance (algorithms, platform, compilers, etc.) from the physics modeling aspects. Thus allowing practitioners to collect runtime information at scale, without the complexities of an actual application, to identifying potential performance trade-offs during the overall co-design process.

In this paper, we evaluate the use of the Multipurpose, Application Centric, Scalable I/O (MACSio)~\cite{MACSio} framework as a lightweight proxy solution to model I/O characteristics of AMR simulations based on the well-established AMReX framework for massively parallel, block-structured AMR applications~\cite{AMReX_JOSS}. We build a simple ``generalized" model based on data collected from AMReX-based inputs and analysis data outputs that can be translated to MACSio command-line parameters in order to replicate data production patterns observed in a simulation. The outline of this paper is as follows: Section~\ref{sec:Background} provides background information on related work and the efforts regarding I/O understanding and characterization. Section~\ref{sec:Methodology} presents the proposed methodology, starting with a brief description of typical inputs and outputs in AMReX-based codes. Generated data is obtained through a series of runs using the Castro simulation framework~\cite{Castro2010} which is included in the AMReX suite of astrophysical hydrodynamics codes. The collected analysis data from AMReX Castro is used to determine the influence of user input parameters, such as: the number of cells, number of mesh levels, output frequency, and the Courant–Friedrichs–Lewy (CFL)~\cite{10.5555/2430727} condition; on the domain decomposition strategy and data generation at different mesh levels (per-task, per-level, and per-timestep). A simple analytical model is then formulated to translate AMReX use-case inputs into arguments to the MACSio executable to model the default I/O behavior of N-to-N output files (1 file per task) at each time step of the simulation. Section~\ref{sec:Results} presents experimental results from the AMReX Castro~\cite{Castro2010} solution of the 2D Sedov blast wave standard hydrodynamics test problem~\cite{Sedov} using multiple configurations running on the Summit supercomputer~\cite{Summit}. Conclusions and future efforts are presented in section~\ref{sec:Conclusion}. Last but not least, the list of public GitLab code repositories with the job scripts and artifact description used to generate data on Summit for this study are presented in the Artifact Description (AD) Appendix~\ref{ap1: Code Repositories} in order to provide the required reproducibility of results presented in this study.
Overall, we are attempting to close the gap between the I/O characterization of specific use-cases in AMReX-based applications and using MACSio as a general, lightweight, kernel proxy I/O tool that can capture the non-linearities of the observed AMR data generation. Having a calibrated lightweight proxy application for I/O is a valuable tool in the codesign process for understanding I/O system dynamic characteristics and trade-offs across AMReX-based codes on exascale systems without having to undertake the deployment of full-scale applications. 

\section{Background}
\label{sec:Background}
Understanding the I/O behavior of codes on supercomputers has long been a research topic that has been receiving greater attention in recent decades as the I/O demands and bottlenecks of scientific applications have become significantly complex. 
In the early 1990s, Akella and Siewiorek~\cite{10.1145/115952.115991} modeled and measured the impact of the I/O bottleneck on system performance.
Around the same time, Miller and Katz~\cite{5348887,160208} characterized the I/O access patterns of several real applications to optimize the use of supercomputer systems. Their performance analysis considered the impact of existing I/O hardware hierarchy of caches, solid-state drives (SSD), disks and tape storage on the very ``bursty" I/O request pattern in which a period of CPU activity was followed by intense I/O activity. Pasquale and Pasquale and Polyzos~\cite{1263486,344330} identified several scientific applications that have predictable ``static" I/O requirements that correlate with CPU activity in production workloads, while also analyzing the ``dynamic" aspect associated with the system. Crandall {\it et al.}~\cite{1383196} studied a wide variety of temporal and spatial write/read patterns in three parallel scalable scientific applications. They found that I/O becomes a major bottleneck due to a combination of inefficient single system policy for a wide diversity in access patterns and request sizes. In the past decade, Carns {\it et al.}~\cite{5937212} proposed a methodology for continuous, scalable I/O characterization by collecting actual application data to help devise strategies that correlate with I/O performance on a Teraflop system. They also presented a tuning case study for an AMR dataset, while confirming the I/O ``burstiness" across several scientific applications. Snyder {\it et al.}~\cite{10.1145/2832087.2832091} discussed different workload generation methods for the abstraction of the I/O. Whereas in recent years, Bez {et al.}~\cite{8924190} incorporated machine learning (ML) algorithms to detect application I/O patterns at runtime.

Proxy applications can be traced early in the development of computing systems as described by De Meis and Weizer~\cite{10.1145/800195.805933} as ``pseudo-programs" whose behavior depend on variable parameters. Proxy applications have also been used to characterize and replicate I/O requirements of large scale scientific codes on supercomputing systems~\cite{7836562}. Recent efforts to characterize I/O via proxy application tools lead to the development of independent ``kernel" tools such as IOR~\cite{5222721}, MACSio~\cite{MACSio}, and Skel~\cite{6130712,8048970}, which abstract out enough characteristics from the application to leverage the use of existing parallel I/O libraries such as HDF5~\cite{SurenByna145} and ADIOS~\cite{GODOY2020100561}. As stated by Dickson {\it et al.}~\cite{7836562}, a ``lightweight" approach is a worthwhile trade-off between manual descriptors and in depth tracing. However, Shan {\it et al.}~\cite{5222721} argue that synthetic I/O benchmarks offer metrics that related directly to system or hardware components, but are extraordinarily difficult to relate back to application requirements. Therefore there is a need to study relationship between full applications I/O requirements and the parameters of a flexible synthetic benchmark.

\section{Methodology}
\label{sec:Methodology}
In this section, the methodology used to characterize the output portion of the I/O from parallel runs of an AMR application is presented along with the model scope and assumptions for the translation to a proxy application using MACSio input parameters. The steps are similar to those identified by Dickson {\it et al.}~\cite{7836562} in their work with MACSio and can be summarized as follows:

\begin{enumerate}
\item Run existing simulations as a point of reference for the I/O generated by the AMReX Castro infrastructure
\item Identify the important input parameters that drive the I/O simulation outputs (checkpoint and analysis data) to determine the basis for a variability study
\item Characterize the output generated given a variety of input parameters in AMReX Castro's configuration file 
\item Evaluate the I/O behavior of AMReX Castro and determine what functionality (if any) is missing from MACSio in order to replicate this behavior
\end{enumerate}

\begin{figure}[ht]
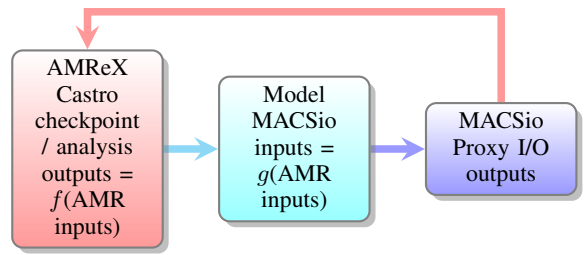

\begin{center}
\smartdiagram[flow diagram:horizontal]{AMReX Castro checkpoint / analysis outputs = $f$(AMR inputs), Model MACSio inputs = $g$(AMR inputs), MACSio Proxy I/O outputs}
\caption{High level flow description of the methodology used in this study to generate a MACSio Proxy I/O model application that captures AMReX based I/O. $g$ represents the functional form of the proposed model based on MACSio functionality.}
\label{fig:overview}
\end{center}
\end{figure}

The flow is illustrated in Fig.~\ref{fig:overview}. This shows a schematic representation of how AMReX Castro inputs and the generated output can be associated with a proxy I/O model. MACSio is used to simulate potential outputs using a set of ``AMR inputs" typically given in a user configuration file. Since the goal is to first understand the role of I/O in an AMR simulation rather than the complexity of the simulation and/or computation, we chose to study the Sedov 2D cylindrical case in Cartesian coordinates for a typical input file\footnote{\url{https://github.com/AMReX-Astro/Castro/blob/main/Exec/hydro_tests/Sedov/inputs.2d.cyl_in_cartcoords}}. This test is readily available in the Castro suite of examples and shows a physical symmetry and can be used to isolate the AMR effects on the I/O on a simple problem.

\subsection{AMReX Castro Parameterized Runs}
\begin{figure}
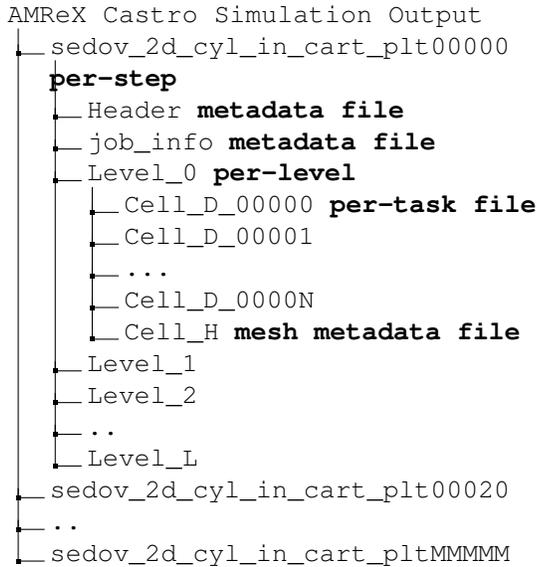

\dirtree{%
.1 AMReX Castro Simulation Output.
.2 sedov\_2d\_cyl\_in\_cart\_plt00000 \textbf{per-step}.
.3 Header \textbf{metadata file}.
.3 job\_info \textbf{metadata file}.
.3 Level\_0 \textbf{per-level}.
.4 Cell\_D\_00000 \textbf{per-task file}.
.4 Cell\_D\_00001.
.4 ....
.4 Cell\_D\_0000N.
.4 Cell\_H \textbf{mesh metadata file}.
.3 Level\_1.
.3 Level\_2.
.3 ...
.3 Level\_L.
.2 sedov\_2d\_cyl\_in\_cart\_plt00020.
.2 ...
.2 sedov\_2d\_cyl\_in\_cart\_pltMMMMM.
}
\caption{Castro file plot analysis output structure for the Sedov 2D cylinder in Cartesian coordinates case.}
\label{fig:file_structure}
\end{figure}

The output directory structure for the generated analysis data (in the form of plot files) can be seen in Fig.~\ref{fig:file_structure}. This default output is done using a $N-to-N$ pattern,  where each of $N$ message passing interface (MPI) tasks writes to a separate file, for the data produced at each refinement level of the simulation. Additional metadata is also produced at the top level in a file called \texttt{Header} and in each directory level in files called \texttt{Cell\_H}. Note that a file is only produced if there is data generated on a particular task at the corresponding mesh level. AMReX also supports the generation of checkpoint-restart data in a similar manner, but we focused on only the plot files for this particular study.

In order to understand the output behavior it is necessary to generate multiple runs of Castro with varying input file configurations. We performed this parameter analysis on Summit using MPI for parallelization. We used the input configuration file in Listing~\ref{lst:input_file}, Appendix~\ref{ap1:config_file} as our baseline to understand the structure and size of the resulting output and to determine the input parameters that influence the output generation. Table~\ref{tab:parameters} shows the input parameters that we focused on for this study.

\begin{table}
\centering
\begin{tabular}{l l}
\hline
  \texttt{amr.max\_step} & maximum expected number of steps\\
  \texttt{amr.n\_cell} &  number of cells at Level 0 in each direction\\
  \texttt{amr.max\_level} &  maximum level of refinement allowed\\
  \texttt{amr.plot\_int} & frequency of plot outputs\\
  \texttt{castro.cfl} & CFL condition\\
 \hline
\end{tabular}
\caption{Subset of AMReX Castro input configuration file parameters varied to understand output behavior in the Sedov hydrodynamics baseline case. }
\label{tab:parameters}  
\end{table}

Aside from those input parameters expected to have direct influence on the analysis data generation in Fig.~\ref{fig:file_structure}, such as the frequency of output file generation, the number of cells at the base mesh (Level 0), and the maximum number of steps, we also focused on parameters that might impact on the mesh refinement. The two primary candidates for these were the CFL condition and the maximum number of refinement levels allowed, so were also considered in the set of parameterized runs.

Each run generated corresponding data that was used to quantify the cumulative output sizes at each requested time interval, refinement level, and task, as described in the hierarchy shown in Fig.~\ref{fig:file_structure}. The selected granularity made it easier to understand the data production at the lowest single file level, while also giving an idea of how balanced (or not) the output was in a simple AMReX application. The results are discussed in more detail in section~\ref{sec:Results}.

\subsection{Modeling outputs with MACSio as a Proxy}
After characterizing the AMReX Castro I/O behavior, the next step was to provide a simple way to simulate the observed I/O patterns using a proxy application. MACSio is a simple and versatile I/O characterization tool designed for HPC workloads. One of the immediate trade-offs when replicating AMReX I/O patterns with MACSio is the level of granularity of the generated output shown in Fig.~\ref{fig:file_structure}.  While AMReX generates outputs based on the tuple $(timestep, refinement\_level, MPI\_task)$, MACSio only provides enough granularity to generate outputs based on $(timestep, MPI\_task)$. Nevertheless, the simplicity of MACSio makes it a worthy candidate for assessing if an approximate solution could simulate deterministic characteristics such as data size, computational overhead, and I/O burstiness at different scales. The latter allows a model to be constructed that could help practitioners understand random and dynamic system characteristics such as bandwidth, file system variability, and scalability, prior to running full AMReX-based simulations on different hardware platforms.

\begin{figure}
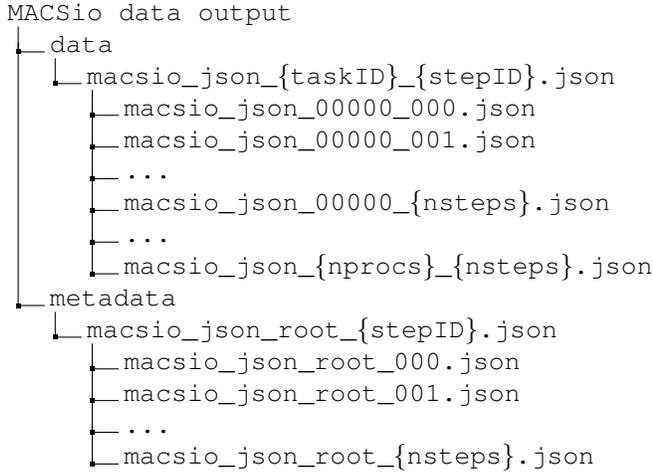

\dirtree{%
.1 MACSio data output.
.2 data.
.3 macsio\_json\_\{taskID\}\_\{stepID\}{.}json.
.4 macsio\_json\_00000\_000{.}json.
.4 macsio\_json\_00000\_001{.}json.
.4 ....
.4 macsio\_json\_00000\_\{nsteps\}{.}json.
.4 ....
.4 macsio\_json\_\{nprocs\}\_\{nsteps\}{.}json.
.2 metadata.
.3 macsio\_json\_root\_\{stepID\}{.}json.
.4 macsio\_json\_root\_000{.}json.
.4 macsio\_json\_root\_001{.}json.
.4 ....
.4 macsio\_json\_root\_\{nsteps\}{.}json.
}
\caption{MACSio's $N-to-N$ output pattern using the \texttt{miftmpl} interface ordered by task and output step, in which \texttt{nsteps} is the total number of steps, and \texttt{nprocs} is the number of MPI tasks.}
\label{fig:macsio_file_structure}
\end{figure}

MACSio provides a simple command-line interface that allows the user to specify how to capture and generate several types of I/O characteristics. We used the default $N-to-N$ output generation from the AMReX-Castro Sedov cases to determine if MACSio can provide a valid approximation to AMReX-Castro I/O patterns for data generation at each timestep. This pattern is shown in Fig.~\ref{fig:macsio_file_structure} for data and metadata files generated from the MACSio executable. 

Table~\ref{tab:macsio_arguments} shows the MACSio  parameters we used in this study to fine-tune the data generation. We found the most important parameters were the \texttt{interface}, \texttt{parallel\_file\_mode}, \texttt{part\_size}, and \texttt{dataset\_growth}.
The latter parameter enables MACSio to approximate non-linear data generation. The ultimate challenge, however, was to fine-tune these parameters to create a proxy I/O model to a level of granularity that is helpful in identifying I/O characteristics in AMRex-Castro.

\begin{table}
\centering
\begin{tabular}{l l}
MACSio Argument & Description \\
\hline
\texttt{interface} & output type hdf5, json (miftmpl), silo \\
\texttt{parallel\_file\_mode} & File Mode: multiple independent, single \\
\texttt{num\_dumps} & number of dumps to marshal (buffer) \\
\texttt{part\_size} & per-task mesh part size \\
\texttt{avg\_num\_parts} & average number of mesh parts per task \\
\texttt{vars\_per\_part} & number of mesh variables on each part \\
\texttt{compute\_time} & rough time between dumps \\
\texttt{meta\_size} & additional metadata size per task \\
\texttt{dataset\_growth} & multiplier factor for data growth \\
\hline
\end{tabular}
\caption{MACSio command line arguments used to model AMReX-Castro outputs.} 
\label{tab:macsio_arguments}  
\end{table}

When comparing the AMReX and Castro inputs in Table~\ref{tab:parameters} and MACSio inputs in Table~\ref{tab:macsio_arguments}, it can be seen that MACSio parameters as such: \texttt{interface} \texttt{parallel\_file\_mode}, and \texttt{num\_dumps} can be easily mapped to the simulation inputs. As such, the constructed model in MACSio will have the functional form shown in Listing~\ref{lst:model}.

\lstset{language=bash,
                basicstyle=\ttfamily\scriptsize,
                showstringspaces=false,
                keywordstyle=\color{blue}\ttfamily,
                stringstyle=\color{red}\ttfamily,
                commentstyle=\color{green}\ttfamily,
                morecomment=[l][\color{magenta}]{\#}
}

\begin{lstlisting}[float,frame=single,mathescape=true,label={lst:model},caption={Proxy app model formulation for mapping MACSio executable to AMReX Castro inputs on Summit.}]
jsrun -n nproc 
macsio
 --interface miftmpl
 --parallel_file_mode MIF nproc
 --num_dumps $\frac{amr.max\_steps}{amr.plot\_int}$
 --part_size $f_{(amr.n\_cell)}$
 --avg_num_parts 1
 --vars_per_part 1
 --compute_time $f_{(platform, all\_inputs)}$
 --meta_size $f_{(all\_inputs)}$
 --dataset_growth $f_{(amr.n\_cell, castro.cfl, amr.max_level, ... )}$
\end{lstlisting}

The remaining challenge then is how to determine the relationship between the \texttt{part\_size} and the \texttt{dataset\_growth} parameters as data is generated from AMR simulations. Other parameters that are ``runtime" in nature such as \texttt{compute\_time} and \texttt{meta\_size} can be determined after collecting data runs. In particular, \texttt{compute\_time} represents a degree of freedom that can be adjusted independently of ``static" data size modeling for ``dynamic" studies to fine-tune the I/O ``burstiness" on a particular platform.

\section{Results}
\label{sec:Results}
This section applies the methodology shown in Section~\ref{sec:Methodology} on parameterized runs on the Castro solution of the 2D Sedov blast wave standard hydrodynamics test problem~\cite{Sedov}. First, data output characteristic are presented for multiple configurations running on Summit. The next step is to construct a functional form of the model shown in Listing~\ref{lst:model} via a minimization process varying the parameters in the MACSio executable. We also list the limitations, current scope and potential use of MACSio as a proxy application to model I/O in AMR simulations.

\subsection{AMReX Castro Sedov Parameterized Outputs}
\begin{figure}[h]
\centering
\subfloat[]{\includegraphics[width=3in]{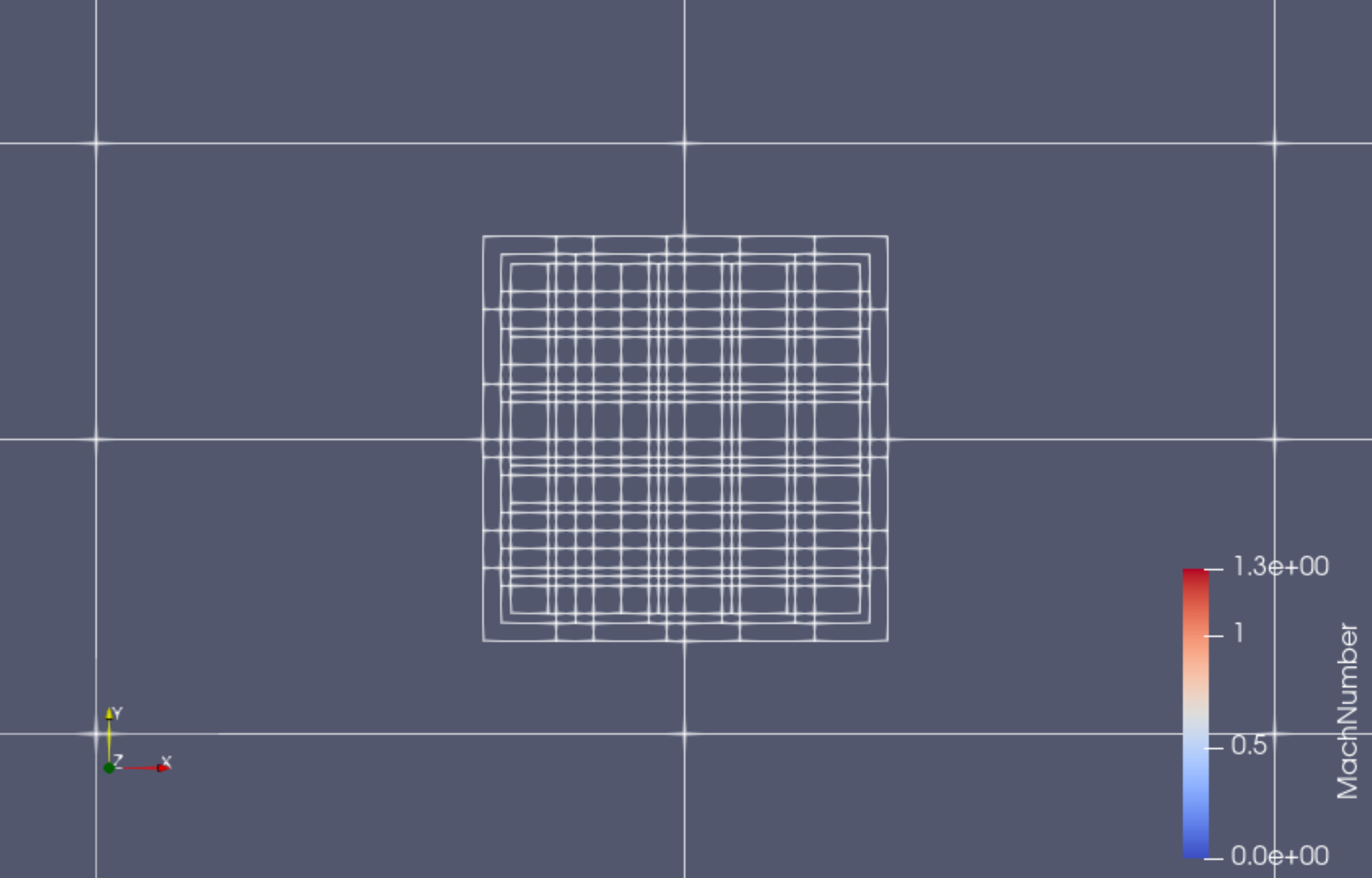}%
\label{fig:sedov_levels}}
\hfil
\hfil
\subfloat[]{\includegraphics[width=3in]{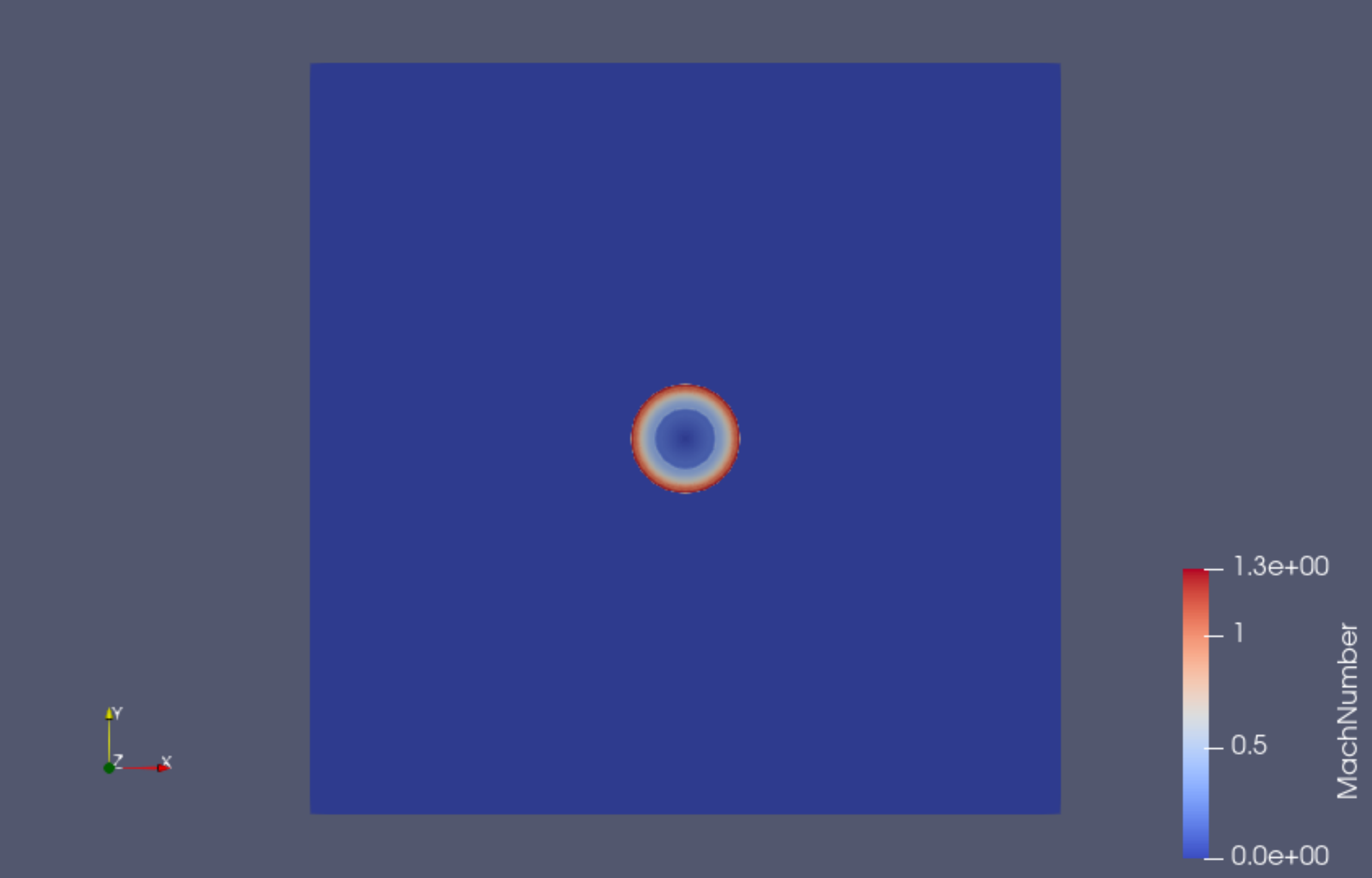}%
\label{fig:sedov_mach}}
\caption{Sedov hydro case 2D cylinder in Cartesian coordinates pivot case used as a benchmark in this study showing (a) AMR mesh showing the moving refined levels, and (b) solution for the Mach number after 20 timesteps. The lower levels follow the solution in the middle affecting overall load balancing, partitioning and I/O.}
\label{fig:sedov}
\end{figure}

The Castro Sedov hydro test generates a straight-forward I/O pattern in which each MPI task outputs the data for each region, at each level, at each requested time interval in the simulation. Based on the resulting outputs, we can infer that this is done in a ``burst buffer" traditional pattern: the computation runs for some time, then the output is generated in a single ``burst" for each time step requested in the input configuration file. Since AMReX provides two methods for writing analysis data, \texttt{WriteSingleLevelPlotfile} and \texttt{WriteMultiLevelPlotfile}, we assumed that the latter is being used by Castro by default based on the file structure shown in Table~\ref{tab:parameters}. Figure~\ref{fig:sedov} illustrates the AMR solution and expected physical results of the baseline symmetric Sedov test. It can be seen that the fine-grained refined levels are generated near the source terms of the hydrodynamics problem as expected in AMR formulations.

In total, 47 runs were performed on Summit at different scales while varying the parameters listed in Table~\ref{tab:parameters} along with the number of files generated as a function of the number of MPI tasks (\texttt{nprocs}). The parameter ranges are summarized in Table~\ref{tab:parameters_Sedov} showing the scope of the present study, along with the \texttt{amr} and \texttt{castro} parameters that were used. The number of MPI tasks (\texttt{nprocs}) and the number of Summit nodes were also varied accordingly to run cases, from small mesh sizes of $32\times32 \approx 1\,K$ cells, to a large mesh size of $131,072\times131,072 \approx 17\,B$ cells using up to 512 Summit nodes, or equivalent to $1/9$ of the 4,608 total system nodes. The expectation is that these ranges would provide enough information to understand the feasibility and scope of MACSio as a simple proxy application by isolating the computational aspects of the baseline AMReX Castro Sedov case. Full configuration scripts are provided for reproducibility purposes as described in Appendix~\ref{ap1: Code Repositories}.

\begin{table}
\centering
\begin{tabular}{l l}
\hline
  Parameter & Range \\
  \texttt{ amr.max\_step} &  40 - 1000 \\
  \texttt{ amr.n\_cell} &  ($32\times32$) - ($131,072\times131,072$)\\
  \texttt{ amr.max\_level} &  2 - 4 (1 to 3 levels) \\
  \texttt{ amr.plot\_int} &  1 - 20 \\
  \texttt{ castro.cfl} & 0.3 - 0.6 \\
  \texttt{ nprocs } &  1 - 1,024 \\
  \texttt{ Summit nodes } &  1 - 512 (1/9 total system) \\
 \hline
\end{tabular}
\caption{AMReX Castro input configuration file parameters range for the Sedov case running imulations to produce different output sizes. }
\label{tab:parameters_Sedov}  
\end{table}

To illustrate the non-regular output characteristics of the generated runs, the independent variable $x$ is considered a function of the user-prescribed number of cells at the ``L0" base level, \texttt{amr.ncells} in Table~\ref{tab:parameters}, and the count of the number of output events up to the maximum number of steps, \texttt{amr.max\_step} in Table~\ref{tab:parameters}, thus resulting in a cumulative quantity. The rationale is that any data output sizes must be proportional to the size of the problem at each requested output event. As a result, the independent variable in our model is expressed as a function of the cumulative independent variable ($x$), and the dependent output size ($y$) at three hierarchical levels:

\begin{eqnarray}
x = output\_counter \times ncells \label{eqn:x}\\
output\_counter = 1, ..., max\_step \nonumber \\
ncells = Nx\,Ny \nonumber \\ 
\nonumber \\
y = data\_output_i \label{eqn:y} \\
i = \texttt{time step, level, task} \nonumber
\end{eqnarray} 

\begin{figure}[h]
\centering
\includegraphics[height=3.7in, width=3.55in]{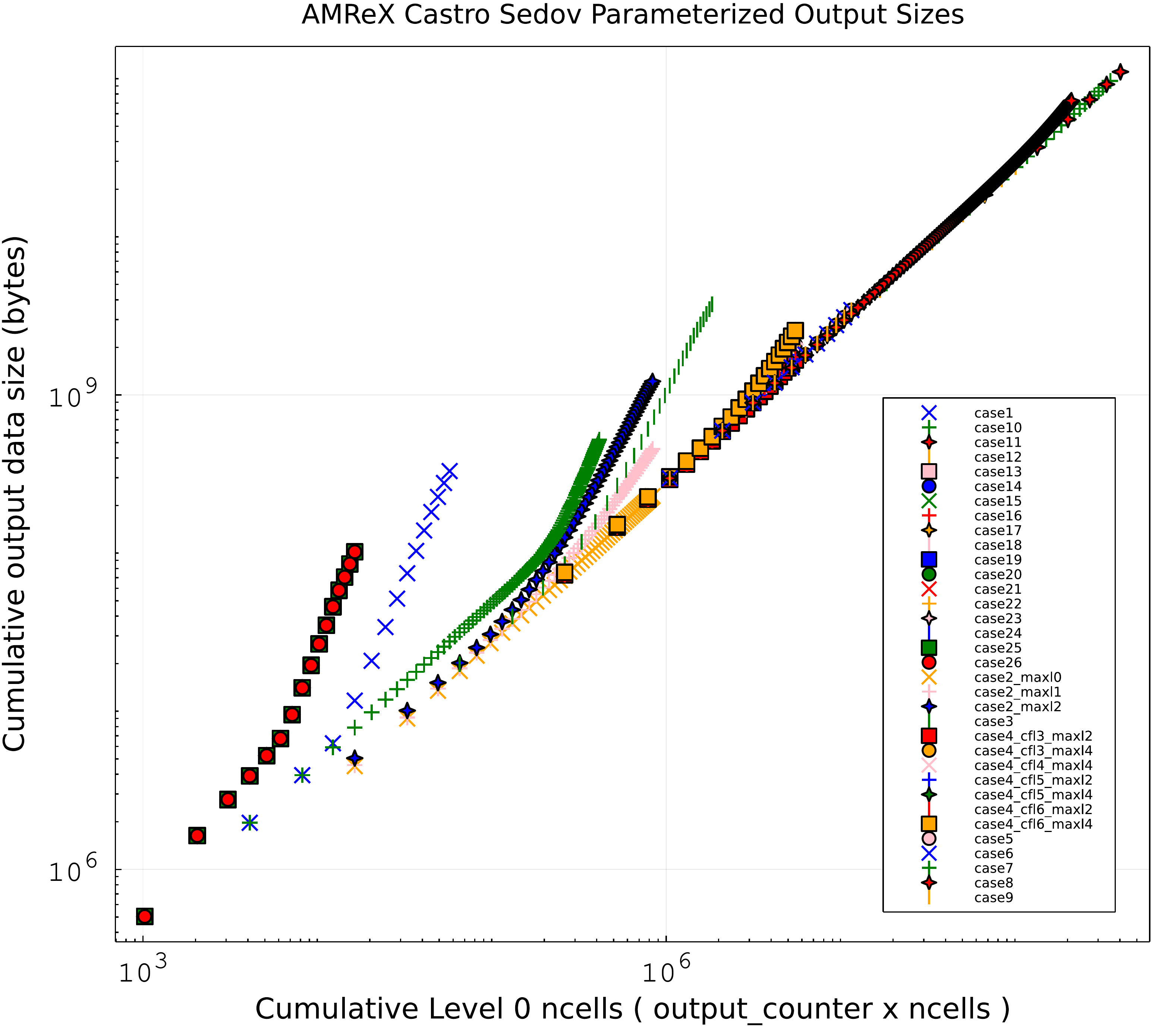}
\caption{Cumulative output size per output step as a function of the cumulative number of output cells as defined in Eqs. (\ref{eqn:x}) and (\ref{eqn:y}) for the Sedov 2D case running on Summit.}
\label{fig:all_cases}
\end{figure}

The cumulative output sizes at each time step are shown in Fig.~\ref{fig:all_cases} for a subset of cases varying the parameters listed in Table~\ref{tab:parameters}. It can be seen the mixed linear and non-linear outputs characteristics for the selected range, while some of the larger cases are excluded for illustration purposes. It can be seen that several runs follow a near-linear trend as expected when the $x$ variable in Eq.~(\ref{eqn:x}) grows with the L0 number of cells. However there is clearly another set of runs that deviate from this linear behavior, and this prompts further inspection in order to better understanding the nature of the generated output.

\begin{figure}[!t]
\centering
\includegraphics[height=3.1in]{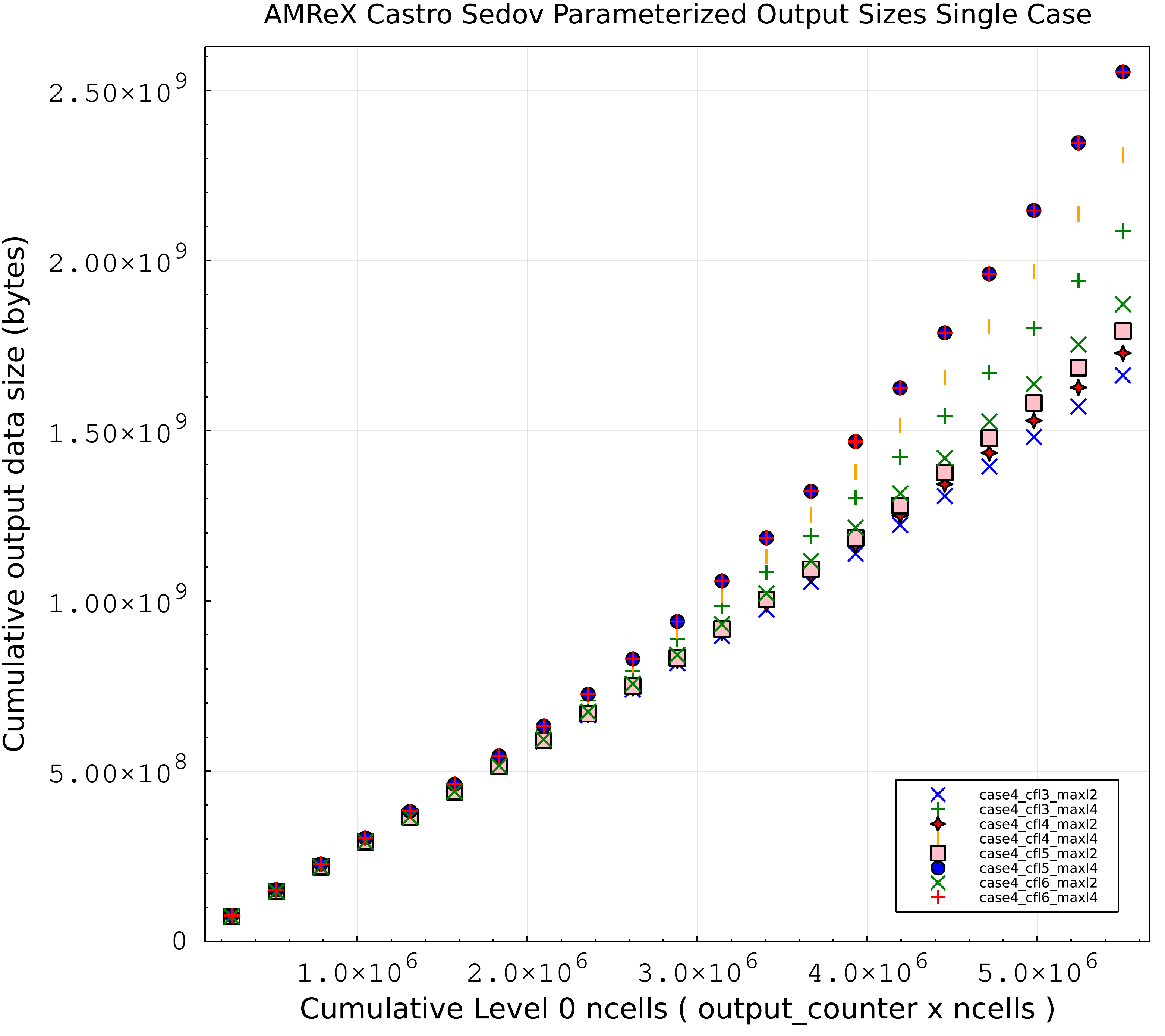}
\caption{Dependency on the CFL number and AMR number of discretization levels for the cumulative output size for a single Sedov simulation run using 2 Summit nodes, 32 tasks and a L0 base mesh of $512\times512$ cells.}
\label{fig:case4}
\end{figure}

To understand the non-linear behavior in output sizes, one of the cases was selected as a pivot (case4) with $N_x=512$ and $N_y=512$ containing 20 outputs. As shown in Fig.~\ref{fig:case4}, it is observed that while the CFL number has some influence on the overall output size, the number of AMR levels has a larger effect and explains the behavior on the other cases that deviate from the observed linear trend in Fig.~\ref{fig:all_cases}. 

\begin{figure}[!htb]
\centering
\includegraphics[height=3.1in]{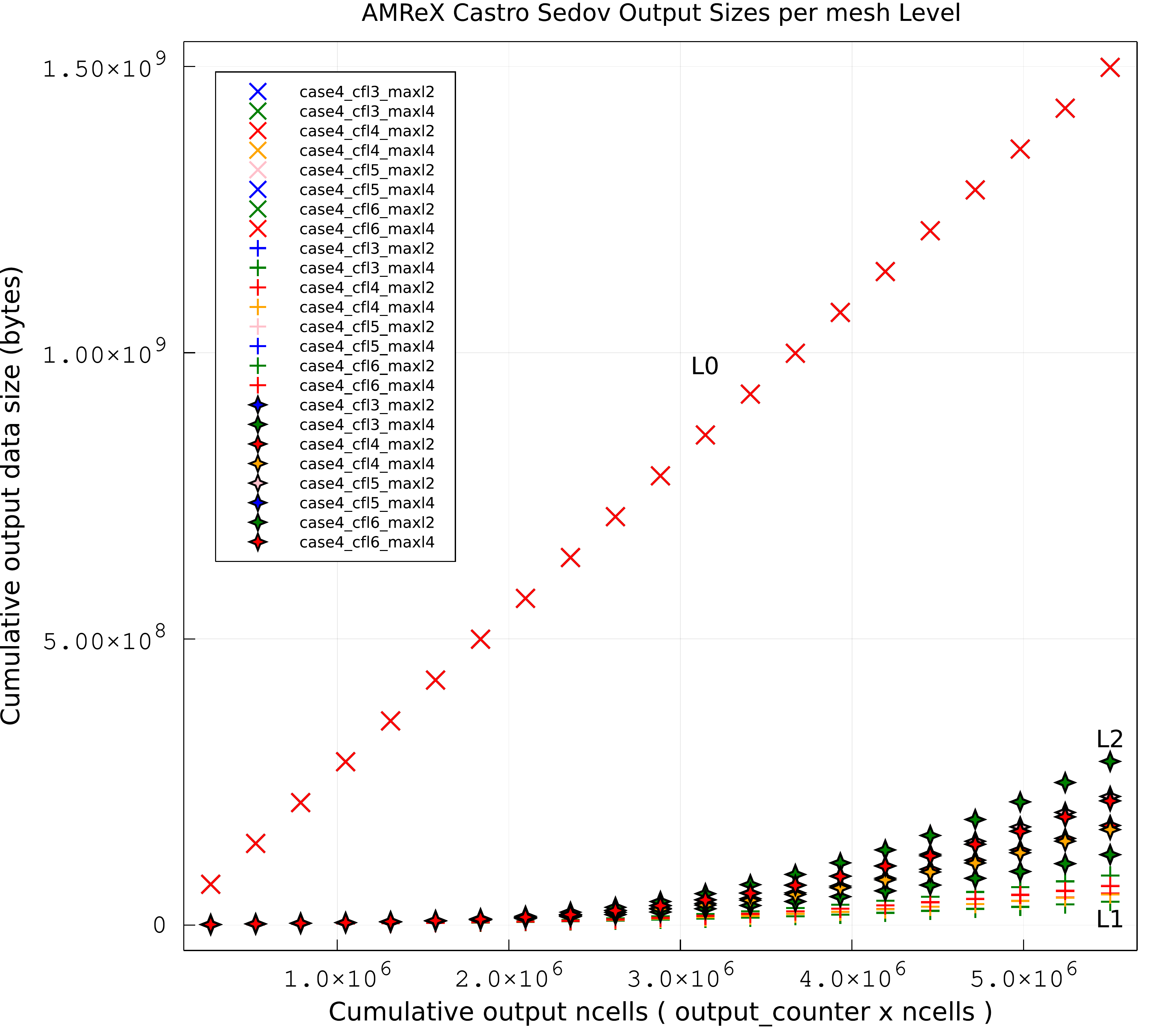}
\caption{Dependency of the cumulative output size for each AMR level (L0,L1,L2) as a function of the cumulative number of output cells and CFL number for the Sedov 2D case.}
\label{fig:case4_levels}
\end{figure}

Further analysis is thus needed to understand how output is distributed among AMR levels, for each level in the output hierarchy $(timestep, level, task)$ as illustrated in Fig.~\ref{fig:file_structure}. The latter is shown in Fig.~\ref{fig:case4_levels} for the pivot case (case4). As expected, the L0 level remains almost constant as it is mainly a function of the user-input number of cells. Subsequent levels in the AMR refinement (L1, L2) are more sensitive as they are driven by the physics ({\it e.g.} larger gradients) and the stability of the meshing algorithm ({\it e.g.} CFL number). An interesting aspect is that the overall ``per-level" output shows a smooth variation. This opens the potential for creating a ``kernel" solution like MACSio, by separating and superimposing the ``linear" behavior from L0 and the ``non-linear" smooth behavior of more refined levels (L1,L2).

\begin{figure}[!htb]
\centering
\includegraphics[width=3.5in,height=3.8in]{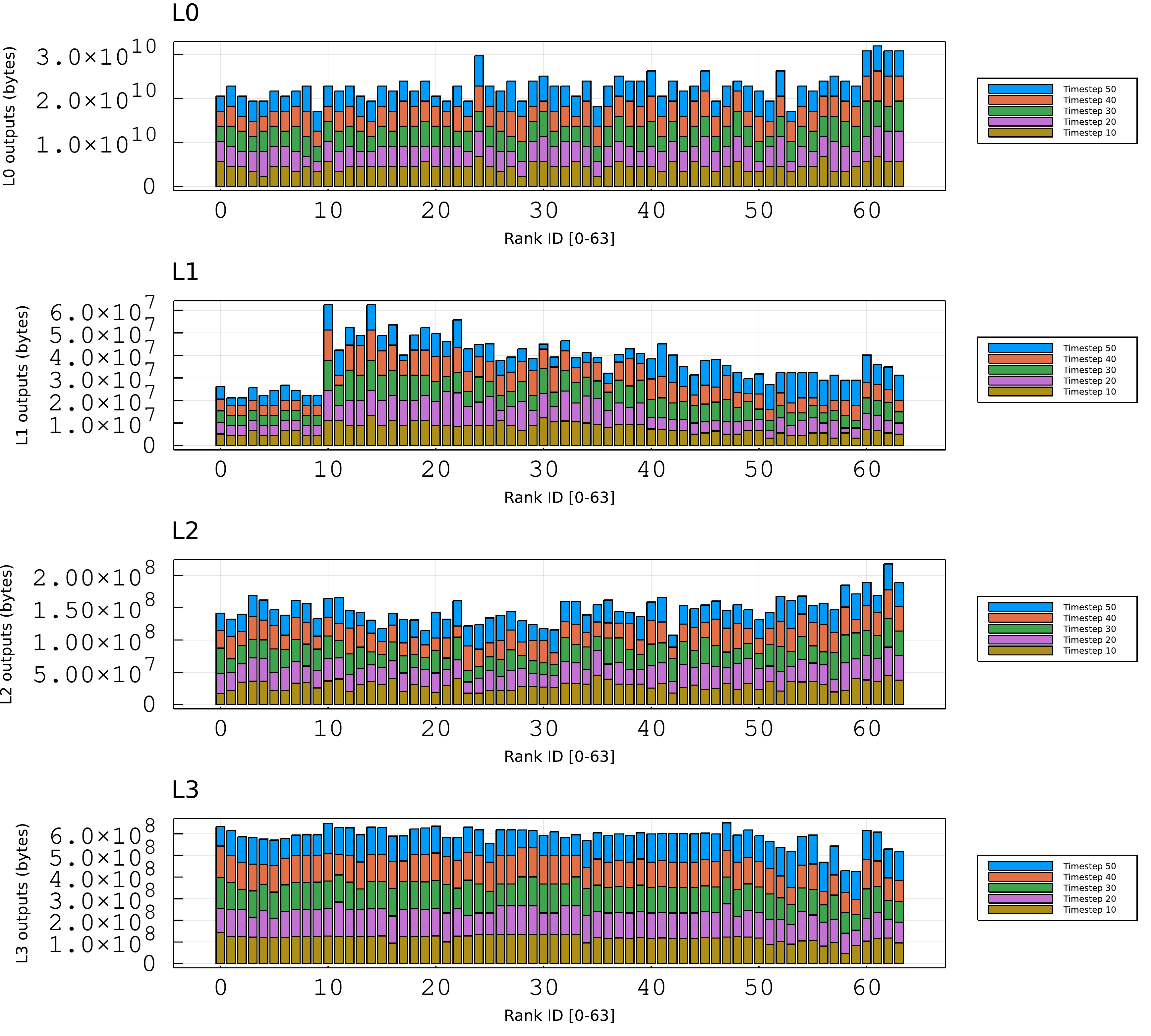}
\caption{Output generation at each timestep per compute task (taskID) for 4 mesh levels in case27, for $1,024\times1,024$ L0 mesh for the AMReX Castro Sedov baseline.}
\label{fig:case27_level_rank}
\end{figure}

Data generation as a function of each mesh level and task is shown in Fig.~\ref{fig:case27_level_rank}. The data produced by each MPI task gives an indication of the AMR effects on load balancing and the predictability of the I/O using a ``kernel" tool like MACSio. Data is generated for 5 output steps, for a simulation of the Sedov case (identified as case27) using 64 ranks on a $1,024\times1,024$ L0 base mesh. As can be seen, AMR effects result in unbalanced loads at all 4 levels of the resulting mesh hierarchy. Further investigation is needed to understand the relationship within AMR levels and MPI decomposition algorithms in AMReX and possible predictability, even in simple configurations such as the Sedov case. Nevertheless, this is an indication that this level of granularity is highly volatile, even for a simple and symmetric problem. Therefore, the model construction using the current MACSio ``kernel" characteristics can only simulate data output loads up to a mesh ``level", but not at the ``rank" level in the AMReX output. This current MACSio limitation is a consequence of its data output model described in Fig.~\ref{fig:file_structure}.

\subsection{MACSio Model}
The next step is to provide a simple way to simulate the observed I/O patterns using a proxy application. As explained in Section~\ref{sec:Methodology}, MACSio was used due to its command-line simplicity and versatility for high-performance simulations. 
While AMReX applications generate an I/O pattern that is dependent on timestep, refinement level, and output type, our goal was to demonstrate that MACSio could capture enough of the non-linearity behavior observed in Fig.~\ref{fig:all_cases} to provide an approximate solution that could simulate deterministic characteristics such as data size, computational overhead, and I/O patterns loads.

\begin{figure}[!htb]
\centering
\includegraphics[width=3.4in]{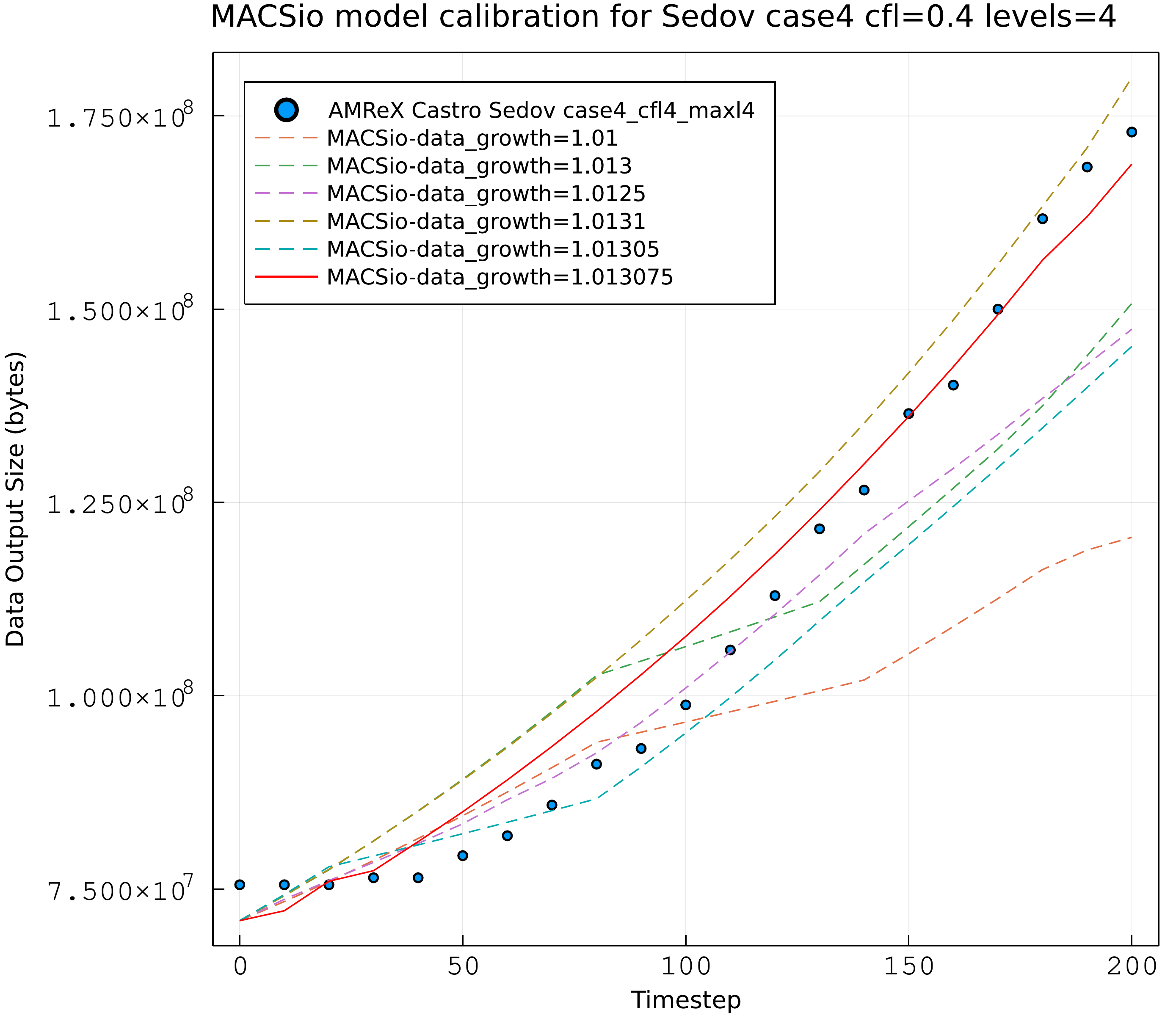}
\caption{Modeling calibration for timestep outputs for the Sedov case4, $cfl = 0.4$, 4 AMR levels using MACSio non-linear kernel approach. Each curve represents a step in the convergence calibration.}
\label{fig:macsio_calibration}
\end{figure}

The AMReX Castro Sedov result (identified as case4) illustrated in Fig.~\ref{fig:case4} was selected as a baseline. The particular case for which the $cfl=0.4$ using 4 levels was compared against several simulations using MACSio for which the \texttt{data\_growth} parameter was calibrated to obtain a non-linear kernel trace for the generated output described in Fig.~\ref{fig:macsio_file_structure}. The initial data size was calibrated against the simulated ``expected" output size multiplied by a correction factor due to its approximate nature in MACSio as a result of constraints involved in creating a valid mesh topology. As a result, a first order approximation for MACSio's \texttt{part\_size} factor in Listing~\ref{lst:model} was estimated as:

\begin{eqnarray}
\texttt{part\_size} = f \frac{8\,Nx\,Ny}{nprocs}\,\,\,\,[bytes] \label{eqn:part_size}\\
f \approx [23-25] \nonumber
\end{eqnarray}

where $f$ is a correction factor due to the difference in nature of the MACSio json-based output and AMReX output file formats. The value of $8$ accounts for the extra bytes in the double precision setup of the Castro executable. The results show that the empirical factor $f$ for the Sedov cases is somewhere around $23$ and $25$, although this value might need to be reevaluated if the number of output fields or a different set of problems is used. Selecting a precise value for $f$ in Eq.~(\ref{eqn:part_size}) depends on the focus of the approximation as a variational problem with two parameters. The latter is illustrated in Fig.~\ref{fig:macsio_calibration} for which the ``best" \texttt{data\_growth} factor is optimized. It can be seen that keeping the initial data size in Eq.~(\ref{eqn:part_size}) fixed would lead to a single parameter optimization problem, which after a few runs shows that MACSio can provide a ``kernel" approximation that is ``close enough" to the output sizes generated with the AMReX Castro Sedov case.
The final solution for $\texttt{data\_growth}=1.013075$ initially deviates from the simulation output sizes, however it becomes close to the correct value as the value as time steps increase, thus providing enough non-linear effect to model a similar behavior. As a result, if the intention is to model data workloads at each time step of the simulation, MACSio provides a simple interface to simulate ``static" loads that can be a starting point for ``dynamic" studies of more random system behaviors.

\begin{figure}[!htb]
\centering
\includegraphics[width=3.4in,height=3.7in]{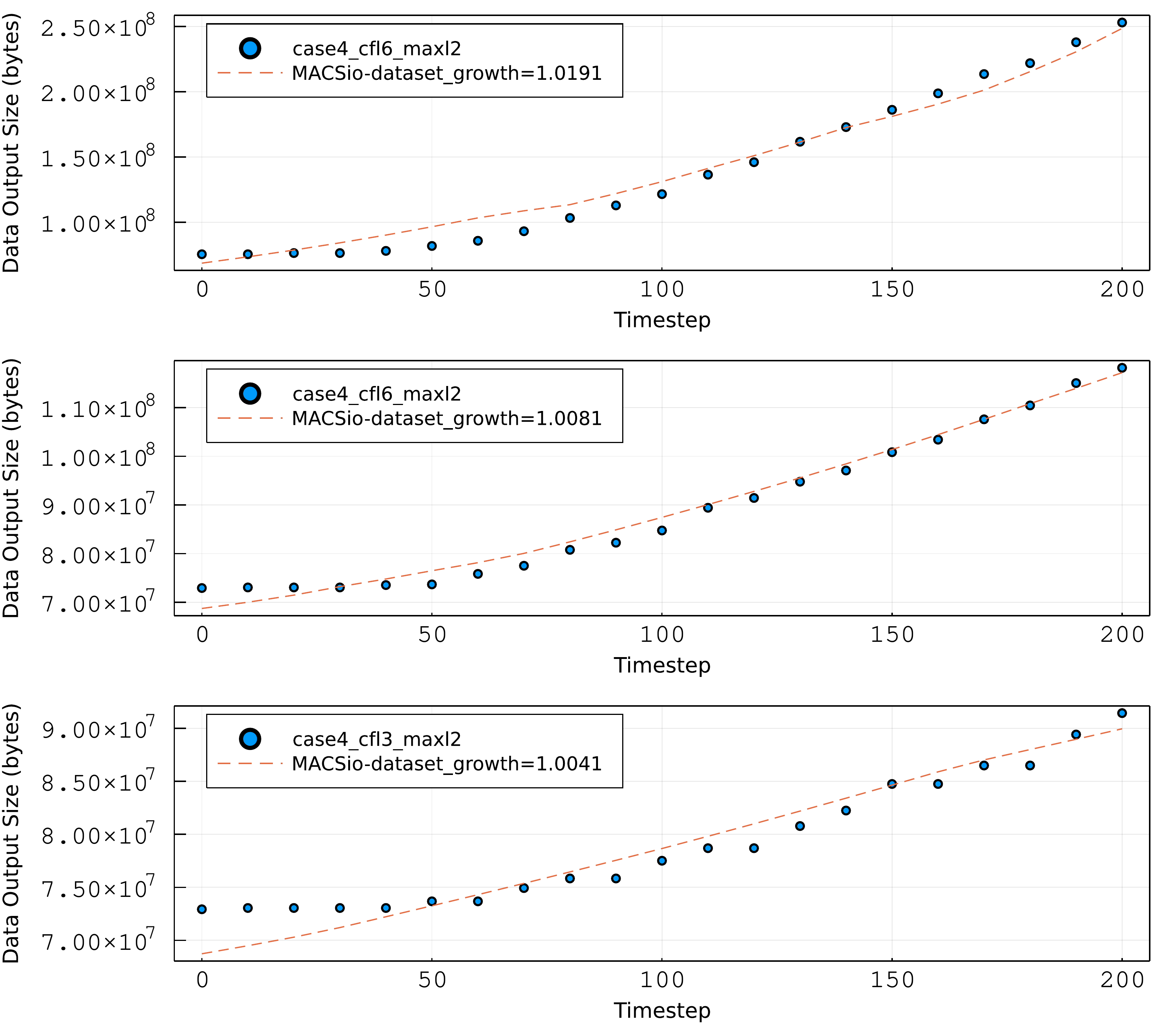}
\caption{Comparison of the baseline Sedov case4 simulation outputs at each time step for different CFL numbers 0.3 (cfl3) and 0.6 (cfl6) and maximum number of mesh levels (maxl=2,4) against the proposed MACSio model.}
\label{fig:macsio_validation}
\end{figure}

The proposed model in Eq.~(\ref{eqn:part_size}) is evaluated for the different cases shown in Fig.~\ref{fig:all_cases} by running MACSio several times while fine tuning the \texttt{data\_growth} parameter to approximate the behavior of the measured data outputs from the Sedov case. To illustrate the validity of the MACSio model, Fig.~\ref{fig:macsio_validation} shows the resulting output sizes from the pivot simulation Sedov case (case4) and the comparison against the proposed MACsio model. It can be seen that after setting the initial data size from Eq.~(\ref{eqn:part_size}) to a constant value $= 1550000 \approx 23.65 \times 512^2 \times 8 / 32 $ the \texttt{data\_growth} becomes a function of the maximum number of levels and the CFL number. Still, this variation is smooth and choosing a small \texttt{data\_growth} value below 1.02 (or 2\%) based on CFL interpolation from these results, can be a good initial guess if further minimization is required depending on use-case ({\it e.g.} machine learning models, quick I/O evaluation, etc.).

\begin{figure}[!htb]
\centering
\includegraphics[width=3.2in]{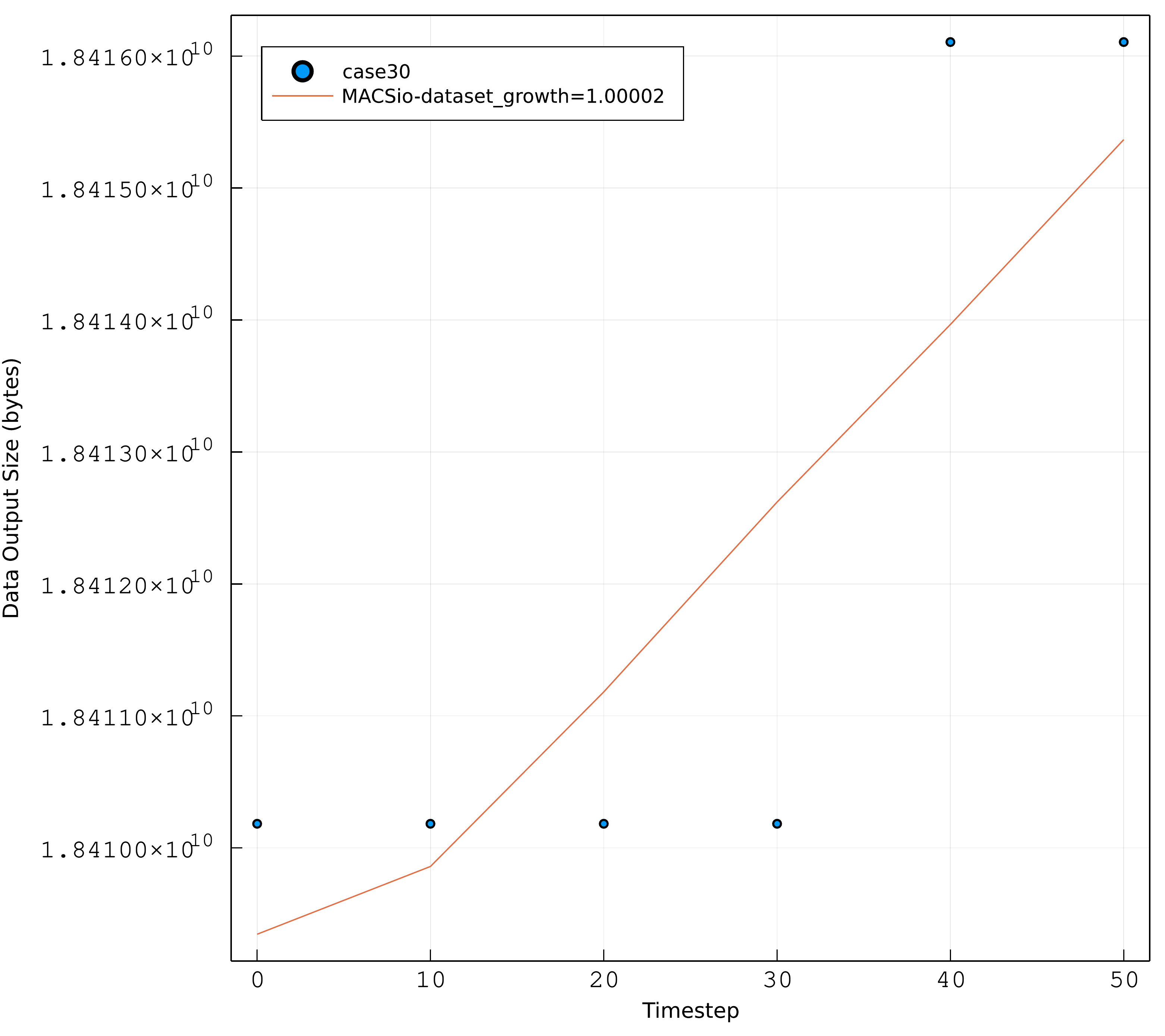}
\caption{Comparison of a large L0 mesh $=8192\times8192$ Sedov non-smooth simulation output against the proposed MACSio kernel model.}
\label{fig:macsio_large_validation}
\end{figure}

Last, but not least we select the large case from Table~\ref{tab:parameters} running on 64 nodes of the Summit supercomputer for a $8192\times8192$ L0 mesh. From Fig.~\ref{fig:all_cases}, it can be seen that as cases become large the non-linearity introduced at the more refined levels becomes less dominant. Nevertheless, while the variation might be less smooth due to a natural reduction in the number of output steps for large scale runs, MACSio can still provide a first-order kernel approximation using the present model. This is illustrated in Fig.~\ref{fig:macsio_large_validation} in which the variation of the output at large scales is less smooth and a sudden jump in output size occurs as convergence of the solution is approached. MACSio can generate kernels that are in the vicinity of these values, while not necessarily providing an exact proxy for the observed non-smooth behavior. We argue that the simplicity of proxy applications, as shown with MACSio for the present study, still provides a reasonable trade-off between complexity and accuracy for modeling desired output workload characteristic at different scales for AMReX based simulations.

\section{Conclusions}
\label{sec:Conclusion}
The data output characteristics of a AMReX-based Castro application under a variety of input conditions are analyzed and modeled using MACSio as a potential candidate for a simple ``kernel"-based proxy I/O application. Results show that MACSio, in its current state of outputting a file for each rank and step, can provide proxy I/O capabilities that are able to model ``per-step" and ``homogeneous per-rank" output sizes of the Sedov hydrodynamic baseline case in AMReX-Castro running on the Summit supercomputer at different scales. In addition, a calibration methodology and an analytical model are provided that relate AMReX Castro inputs with those of MACSio, thus resulting in a lightweight proxy approach to conduct initial studies of AMReX I/O modeled characteristics without having to run a full simulation in parameterized studies. The latter simple analytical model becomes useful as storage systems evolve, along with the need to understand the co-design trade-offs when producing data at scale using AMR-based simulations. Furthermore, this simplified proxy ``kernel"-based approach can be a good initial candidate for follow up studies on predictive I/O sizes, as well as dynamic random system characteristics that could potentially benefit from machine-learning approaches as more data becomes available. Since the ultimate goal is to improve the understanding of AMR output generation, proxy applications combined with simple modeling approaches can be a powerful predictive tool for autotuning more complex parallel I/O workload patterns in current pre and upcoming exascale supercomputing platforms. The latter autotuning aspect is something we would like to explore in subsequent studies.

\section*{Acknowledgements}
This research was supported by the Exascale Computing Project (17-SC-20-SC), a collaborative effort of the U.S. Department of Energy Office of Science and the National Nuclear Security Administration. This research used resources of the Oak Ridge Leadership Computing Facility at the Oak Ridge National Laboratory, which is supported by the Office of Science of the U.S. Department of Energy under Contract No. DE-AC05-00OR22725.
This manuscript has been authored by UT-Battelle, LLC, under contract DE-AC05-00OR22725 with the US Department of Energy (DOE). The publisher acknowledges the US government license to provide public access under the DOE Public Access Plan (\url{https://energy.gov/downloads/doe-public-access-plan})

\bibliographystyle{IEEEtran}
\bibliography{IEEEabrv,paper.bib}

\appendices

\section{Artifact Description for Reproducibility}
\label{ap1: Code Repositories}

Reproducibility steps for the current data generated for this study on the Summit~\cite{Summit} supercomputer. Steps include links to the publicly available artifacts used in this study: \\
\begin{outline}[enumerate]
\1 Build AMReX Castro version 20.10.  Sedov case on Summit using gcc-9: 
 \2 Source Code: \\
 \url{https://github.com/AMReX-Astro/Castro} 
 \2 Build Instructions: \\ \url{https://amrex-astro.github.io/Castro/docs/getting_started.html} \\
    
\1 Modify the template for Summit job scripts for the cases generated in this study:
    \2 Job scripts repository: \\
    \url{https://code.ornl.gov/rse-public/amr_io_evaluation} \\
    
\1 Build MACSio v1.1 from source on Summit using gcc-9 (requires json-cwx):
    \2 Source code: \url{https://github.com/LLNL/MACSio} 
    \2 Instructions: \\
    \url{https://macsio.readthedocs.io/en/latest/macsio_building.html#installing-manually} \\
\1 Apply the proposed model in Eq.(\ref{eqn:part_size}) for an initial \texttt{part\_size} in the MACSio executable and \texttt{data\_growth} $\approx$ 1.0 - 1.02 . The greater the cfl and number of levels, the greater the \texttt{data\_growth}. \\

\1 Post-processing tools:
   \2 JupyterHub capability on Summit~\cite{Summit} Notebook: \\ %
   \url{https://code.ornl.gov/rse-public/amr_io_evaluation/-/blob/master/amrex/castro/sedov/MACSio_AMReX_Comparison.ipynb}
   \2 jexio Julia~\cite{Bezanson2017-ca} package: \\
   \url{https://code.ornl.gov/wfg/jexio} \\
\end{outline}

\newpage
\section{AMReX Castro configuration file}
\label{ap1:config_file}

\begin{lstlisting}[frame=single,label={lst:input_file},caption={Input configuration file for the AMReX-Astro Castro Sedov inputs.2d.cyl\_in\_cartcoords case.}]
# INPUTS TO MAIN PROGRAM  
max_step = 500
stop_time = 0.1

# PROBLEM SIZE & GEOMETRY
geometry.is_periodic =  0 0
geometry.coord_sys   =  0  # 0 => cart
geometry.prob_lo     =  0    0
geometry.prob_hi     =  1    1
amr.n_cell           = 32   32

# BC FLAGS 
# 0 = Interior 3 = Symmetry
# 1 = Inflow   4 = SlipWall
# 2 = Outflow  5 = NoSlipWall
castro.lo_bc       =  2   2
castro.hi_bc       =  2   2

# WHICH PHYSICS
castro.do_hydro = 1
castro.do_react = 0

# TIME STEP CONTROL
# CFL number for hyperbolic system
castro.cfl            = 0.5     
# scale back initial timestep
castro.init_shrink    = 0.01
# maximum increase in dt
# over successive steps
castro.change_max = 1.1    

# DIAGNOSTICS & VERBOSITY
# timesteps between computing mass
castro.sum_interval = 1       
# verbosity in Castro.cpp
castro.v = 1       
# verbosity in Amr.cpp
amr.v = 1
# name of grid logging file
#amr.grid_log = grdlog  

# REFINEMENT / REGRIDDING
# maximum level number allowed
amr.max_level = 3       
# refinement ratio
amr.ref_ratio = 2 2 2 2 
# how often to regrid
amr.regrid_int = 2       
# block factor in grid generation
amr.blocking_factor = 8       
amr.max_grid_size = 256

# CHECKPOINT FILES
# root name of checkpoint file
amr.check_file = sedov_2d_cyl_in_cart_chk   
# number of timesteps between checkpoints
amr.check_int   = 20       

# PLOTFILES
# root name of plot (analysis) file
amr.plot_file = sedov_2d_cyl_in_cart_plt
# number of timesteps between plots
amr.plot_int    = 20
amr.derive_plot_vars=ALL

# PROBIN FILENAME
amr.probin_file = 
  probin.2d.cyl_in_cartcoords
\end{lstlisting}

\end{document}